\documentclass{webofc}
\usepackage[varg]{txfonts}   
\usepackage{latexsym,amssymb,amsfonts,amsmath,slashed}
\usepackage{graphicx}

\newcommand{\f}[2]{\frac{#1}{#2}}

\newcommand{\la}{\langle}
\newcommand{\ra}{\rangle}

\newcommand{\Dslash}{\slashed{D}}

\begin{document}

\title{Localisation of Dirac eigenmodes and confinement in gauge
  theories: the Roberge-Weiss transition}

\author{\firstname{Marco}
  \lastname{Cardinali}\inst{1}
  \and \firstname{Massimo}
  \lastname{D'Elia}\inst{1}
  \and \firstname{Francesco}
  \lastname{Garosi}\inst{2}
  \and\firstname{Matteo}
  \lastname{Giordano}\inst{3}\fnsep\thanks{\email{giordano@bodri.elte.hu}}
}

\institute{
  Dipartimento di Fisica dell'Universit\`a di Pisa and INFN
  - Sezione di Pisa, Largo Pontecorvo 3, I-56127 Pisa, Italy
  \and
  SISSA, Via Bonomea 265, I-34136, Trieste, Italy
  \and
  ELTE E\"otv\"os Lor\'and University, Institute for
  Theoretical Physics, P\'azm\'any P\'eter s\'et\'any 1/A, H-1117, Budapest,
  Hungary
}

\abstract{Ample numerical evidence from lattice calculations shows a
  strong connection between the confining properties of gauge theories
  at finite temperature and the localisation properties of the
  low-lying Dirac eigenmodes. In this contribution we discuss recent
  progress on this topic, focussing on results for QCD at imaginary
  chemical potential ${\mu}_I/T = \pi$ at temperatures above the
  Roberge-Weiss transition temperature. These confirm the general
  picture of low modes turning from delocalised to localised at the
  deconfinement transition, in a previously unexplored setup with a
  genuine, physical transition in the presence of dynamical
  fermions. This further supports the use of Dirac eigenmodes as a
  tool to investigate the mechanisms behind confinement and the
  deconfinement transition.}

\maketitle

\section{Introduction}
\label{sec:intro}

While the crossover nature and the location of the QCD transition at
finite temperature have been firmly established, the mechanism that
drives it is still a matter of active research. In particular, one
would like to understand why both confining and chiral properties
change dramatically in the same relatively narrow temperature range,
where quarks and gluons deconfine and chiral symmetry gets
restored. In fact, linear confinement of quarks and spontaneous
breaking of chiral symmetry apply to QCD only in an approximate sense,
originating from opposite quark mass limits, and it is not yet fully
understood how they affect each other.

Quarks are confined by a linear potential only in the ``quenched''
limit of infinite quark mass, where QCD reduces to pure SU(3) gauge
theory. In this setting, deconfinement is signalled by the spontaneous
breaking of the exact centre symmetry of the theory by the ordering of
the local Polyakov loops, resulting in a vanishing string tension. The
Polyakov loop,
\begin{equation}
  \label{eq:pl}
  P(\vec{x}\mkern 2mu) =
  \textrm{Pexp}\left\{ig\int_0^{\f{1}{T}}dt\,A_4(t,\vec{x}\mkern 2mu) \right\}\,,
\end{equation}
is the holonomy of the gauge field around the temporal direction,
which in the imaginary-time formulation of finite-temperature gauge
theories is compactified to a circle of size equal to the inverse of
the temperature, $T$. In pure gauge theory at high temperature the
local Polyakov loops tend to align to one of the elements of the gauge
group centre, $\mathbb{Z}_3$, thus breaking centre symmetry, with the
trivial element being the one selected when pure gauge theory is
approached as the quenched limit of QCD. Deconfinement is then
associated with the ordering of local Polyakov loops, and while to a
small extent this is present in QCD already at low temperatures, it
becomes considerably stronger at the crossover.

On the other hand, chiral symmetry is a symmetry only in the limit of
massless quarks, and only there one can speak of its spontaneous
breaking in a strict sense. In this limit, chiral symmetry is
spontaneously broken to its vector part by a quark condensate,
resulting from the accumulation of eigenvalues of the Dirac operator
$\Dslash$ near the origin. This is made clear by the Banks-Casher
formula~\cite{Banks:1979yr},
\begin{equation}
  \label{eq:BC}
  \Sigma=  |\la \bar{\psi}\psi\ra | = \int_0^\infty d\lambda\,  \rho(\lambda)
  \f{2m}{\lambda^2 + m^2}\,,\qquad  \rho(\lambda)  \equiv
  \lim_{V\to\infty}\f{T}{V}\Big\la \sum_n \delta(\lambda -
  \lambda_n)\Big\ra\,, 
\end{equation}
where $\rho(\lambda)$ is the spectral density of $\Dslash$. Here
angular brackets denote the expectation value in the sense of the
Euclidean functional integral in the imaginary-time formulation of
finite-temperature gauge theories, $i\lambda_n$ are the eigenvalues of
$\Dslash$ in a given gauge configuration, and $V$ is the spatial
volume of the system. In the chiral limit one finds precisely
$\Sigma\to \pi\rho(0^+)$. Even for physical quark masses, the
crossover is characterised by a dramatic reduction of the density of
near-zero modes, and so the approximate restoration of chiral symmetry
is signalled by the depletion of the near-zero spectral region.

Despite their very different origins, deconfinement and chiral
restoration show a close relationship not only in QCD, but also in
other QCD-like gauge theories. In theories with a single, genuine
deconfining phase transition (e.g., pure gauge theory, or lattice QCD
with $N_f=3$ staggered fermions on coarse
lattices~\cite{Karsch:2001nf}), this is associated with a general
improvement of the chiral properties, in particular with a decrease of
the chiral condensate. In theories with two genuine transitions (e.g.,
QCD with $N_f=2$ adjoint fermions~\cite{Karsch:1998qj}), deconfinement
preceeds chiral restoration. This suggests that deconfinement strongly
affects the chiral properties of the theory, although it is not yet
clear by means of what mechanism.

What could play a role in this mechanism is the localisation of the
low-lying Dirac modes taking place at the finite temperature
transition in
QCD~\cite{GarciaGarcia:2006gr,Kovacs:2012zq,Cossu:2016scb,Holicki:2018sms,
  Giordano:2013taa,Nishigaki:2013uya,Ujfalusi:2015nha} and other gauge
theories (see Ref.~\cite{Giordano:2021qav} for a recent review). While
delocalised at low $T$, low Dirac modes become localised above the
transition, up to a critical point in the spectrum known as mobility
edge. Ample numerical evidence shows that the connection between
deconfinement and low-mode localisation is a general feature of gauge
theories. In particular, localised low modes appear precisely at
deconfinement when this takes place through a genuine phase
transition. The general connection between deconfinement and low-mode
localisation is understood qualitatively in terms of the
``sea/islands'' picture of
localisation~\cite{Bruckmann:2011cc,Giordano:2015vla,
  Giordano:2016cjs,Giordano:2016vhx}. In this picture, the ordering of
local Polyakov loops into a ``sea'', that signals deconfinement, is
responsible for the opening of a gap in the spectrum of $\Dslash$,
that can be filled by eigenmodes only at the price of localising on
``islands'' of Polyakov-loop fluctuations. This leads one to expect
the localisation of low Dirac modes in the deconfined phase of a
generic gauge theory, an expectation so far supported by all the
available numerical
investigations~\cite{Giordano:2021qav}.\footnote{The connection
  between localisation and chiral symmetry breaking has received less
  attention in recent years. For the possible consequences of the
  presence of localised modes for chiral symmetry breaking, see
  Refs.~\cite{Giordano:2020twm,Giordano:2022ghy}.}

Until recently, the numerical evidence for a strong
deconfinement/localisation connection, with localised modes appearing
right at the critical temperature, came mostly from pure gauge
theories with a sharp transition. The only exception was the toy model
for QCD provided by $N_f=3$ staggered fermions on coarse
lattices~\cite{Giordano:2016nuu}, where the transition is only a
lattice artefact and does not survive the continuum limit. The first
study in a theory with a genuine and physical phase transition in the
presence of dynamical fermions was done by us in
Ref.~\cite{Cardinali:2021fpu}, looking at the Roberge-Weiss transition
in QCD with an imaginary chemical potential $\mu_I$ at
$\hat{\mu}_I\equiv \mu_I/T=\pi$. Our results confirm the general
expectation, and further support the strong connection between
deconfinement and localisation.

In this contribution, after a brief review of localisation in
Sec.~\ref{sec:loc}, of the sea/islands picture in Sec.~\ref{sec:si},
and of QCD at finite imaginary chemical potential in
Sec.~\ref{sec:RW}, we discuss our results obtained in QCD at
$\hat{\mu}_I=\pi$ above the Roberge-Weiss temperature in
Sec.~\ref{sec:RW_num}.

\section{Localisation} 
\label{sec:loc}

Eigenmode localisation is a concept originating in the study of random
Hamiltonians mimicking disordered systems~\cite{lee1985disordered}.
Qualitatively speaking, the localised modes of a random Hamiltonian
$H$ are those supported essentially only in a finite region of space,
where their amplitude squared
$\Vert \psi(x)\Vert^2 = \sum_{\alpha} \psi_{\alpha
}(x)^*\psi_{\alpha}(x)$ is not negligible,
$\Vert \psi(x)\Vert^2 \sim 1/L_0^d$ with $d$ the spatial dimension and
$L_0$ the linear size of the localisation region.
Here $\sum_\alpha$ denotes the sum over any discrete spacetime or
internal index. Delocalised modes instead extend throughout the
system, with $\Vert \psi(x)\Vert^2 \sim 1/L^d$ in a finite spatial box
of side $L$.\footnote{Modes with
  $\Vert \psi(x)\Vert^2 \sim 1/L^\alpha$, $0<\alpha<d$, are known as
  ``critical'' in the condensed matter literature.}  A more accurate
characterisation is in terms of how the average spatial size of the
modes scales with the system size. This is obtained from their inverse
partecipation ratio (IPR), averaged over modes in the spectral region
of interest and over realisations of $H$,
\begin{equation}
  \label{eq:IPR}
  \overline{\textrm{IPR}}(\lambda,L)  = \f{\la \sum_n
  \delta(\lambda-\lambda_n)\textrm{IPR}_n\ra}{\la \sum_n
  \delta(\lambda-\lambda_n)\ra}\,, 
  \qquad
  \textrm{IPR}_n = \int d^dx \, \Vert
  \psi_n(x)  \Vert^4  
  \,,
\end{equation}
where $\psi_n$ and $\lambda_n$ are the eigenmodes and eigenvalues of
$H$ for some disorder realisation, $H\psi_n= \lambda_n \psi_n$, and
angular brackets indicate the average over disorder. The inverse of
the IPR measures the spatial size of the mode: for localised modes
$\overline{\textrm{IPR}}(\lambda,L)$ remains finite as $L\to\infty$,
while for delocalised modes $\overline{\textrm{IPR}}(\lambda,L)\to 0$.

\begin{figure}[t]
  \centering
  \includegraphics[width=0.53\textwidth,clip]{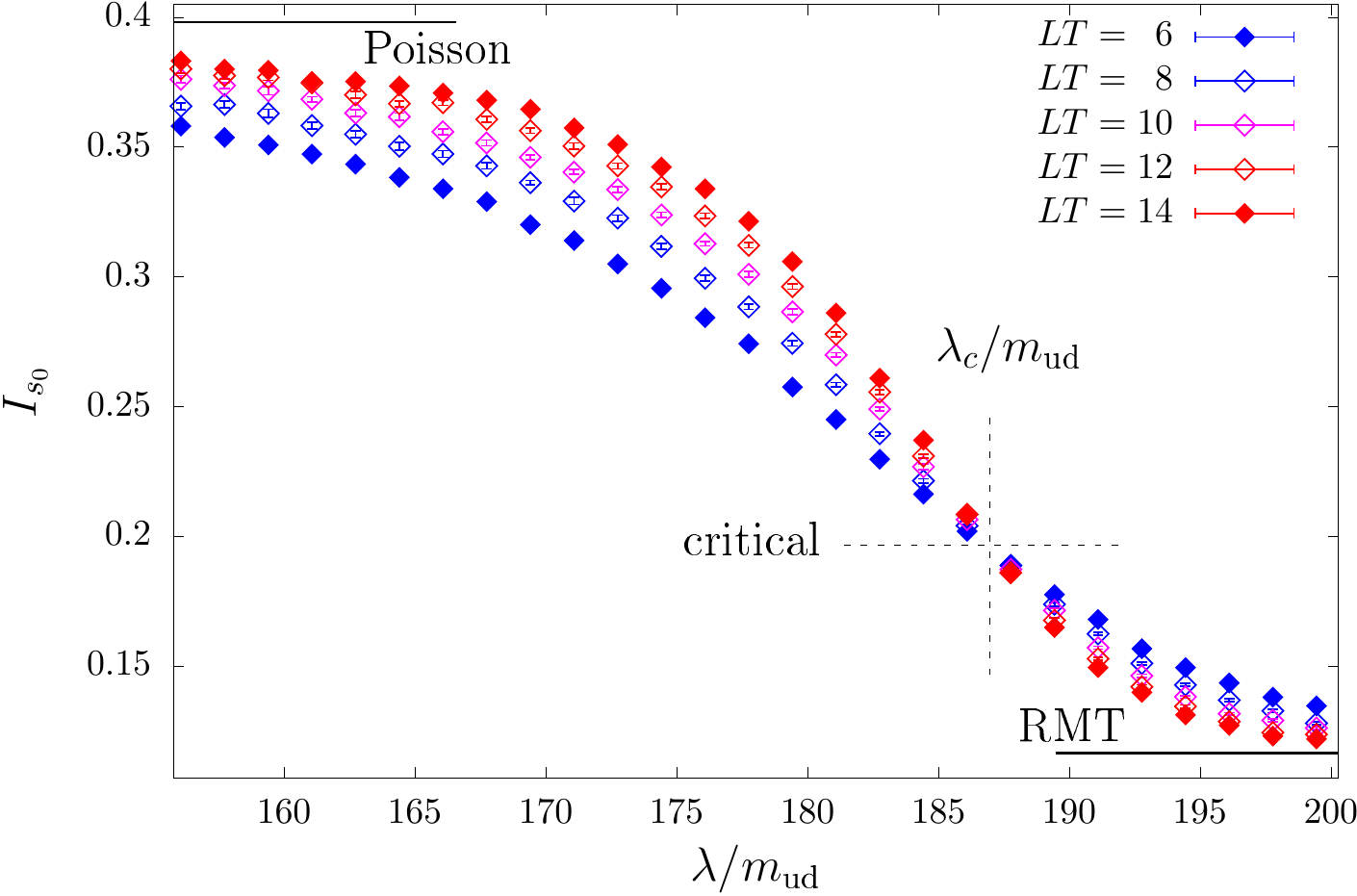}\hfil\hfil
  \raisebox{0.6cm}{\includegraphics[width=0.47\textwidth,clip]{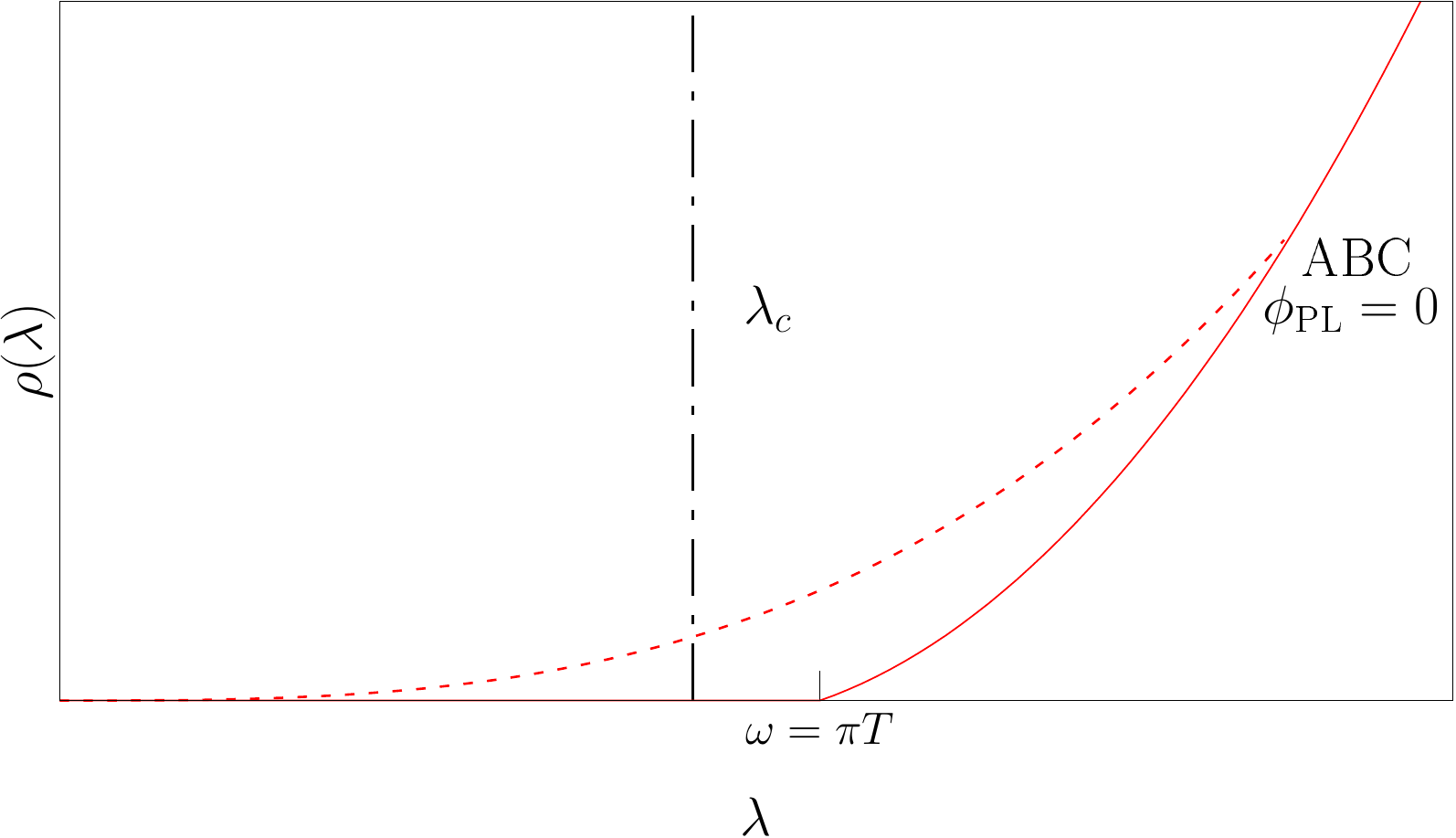}}

  \caption{Left: $I_{s_0}$ [see Eq.~\eqref{eq:iulsd}] along the Dirac
    spectrum (in units of the light-quark mass $m_{\textrm{ud}}$) for
    2+1 flavour lattice QCD at $T=394~\textrm{MeV}$, for several
    spatial sizes $L$ (here the lattice spacing is
    $a=0.125~\textrm{fm}$, and $s_0\simeq 0.508$). The mobility edge,
    the critical value of the statistics, as well as the Poisson and
    RMT expectations are also shown. (Data from
    Ref.~\cite{Giordano:2013taa}.) Right: Sketch of the spectral
    density in high-temperature QCD ($\hat{\mu}_I=0$). The solid line
    is the spectral density in the case of vanishing gauge fields,
    starting at the gap $\omega=\pi T$, while the dashed line
    qualitatively shows the spectral density after the inclusion of
    localised modes penetrating the gap, and $\lambda_c$ indicates the
    position of the mobility edge.}
\label{fig:iulsd}       
\end{figure}

A convenient way to detect localisation is through the study of the
statistical properties of the
eigenvalues~\cite{altshuler1986repulsion}.  Since delocalised modes
are easily mixed by disorder fluctuations, the corresponding
eigenvalues are expected to obey the statistics of the appropriate
ensemble of Random Matrix Theory (RMT)~\cite{mehta2004random},
depending on the symmetry class of $H$. Localised modes, on the other
hand, are sensitive only to disorder fluctuations in their
localisation region, and so the corresponding eigenvalues should
fluctuate independently for uncorrelated disorder, or more generally
for disorder with short-range correlations, therefore obeying Poisson
statistics. A simple way to analyse the statistical properties of the
spectrum is in terms of the distribution $p_\lambda(s)$ of the
unfolded level spacings, computed locally in the spectrum.  Unfolded
level spacings are defined as
\begin{equation}
  \label{eq:uls}
  s_i = \f{\lambda_{i+1}-\lambda_i}{\la
    \lambda_{n+1}-\lambda_n\ra_\lambda}\,,
\end{equation}
where ${\la \lambda_{n+1}-\lambda_n\ra_\lambda}$ is the average level
spacing in the spectral region of interest. The behaviour of
$p_\lambda(s)$ is universal, depending only on the type of statistics
and not on the details of the model, and is explicitly known both for
Poisson and RMT statistics. Localisation can then be detected by
comparing features of $p_\lambda(s)$ to the known expectations. To
this end, a particularly convenient observable is the integrated
unfolded level spacing distribution~\cite{Shklovskii:1993zz}
\begin{equation}
  \label{eq:iulsd}
  I_{s_0} = \int_0^{s_0}d s\,p_\lambda(s)\,,
\end{equation}
where $s_0$ is chosen to maximise the difference between the Poisson
and RMT expectations.

Disordered systems typically display disjoint spectral regions of
localised and delocalised modes separated by mobility edges, where the
localisation length of the localised modes diverges and the system
undergoes a second-order phase transition known as Anderson
transition~\cite{Evers:2008zz}.  As the system size increases,
observables like $I_{s_0}$ tend to their Poisson or RMT expectation
depending on the localisation properties of the modes in the spectral
region of interest, except at a mobility edge, where they are
scale-invariant and take the critical value appropriate for the
symmetry class of the system. This allows for an accurate
determination of the position of the mobility edge via a finite-size
scaling analysis or, if the critical value is known, by looking for
the point where the observable takes it.

The discussion above carries over unchanged to the study of the
eigenmodes of the Dirac operator, which is precisely ($-i$ times) a
random Hamiltonian with disorder provided by the fluctuations of the
gauge fields, over which one integrates in the functional-integral
formulation of the theory. An example of how to detect localisation of
Dirac eigenmodes using spectral statistics is shown in
Fig.~\ref{fig:iulsd} (left), in the case of high-temperature lattice
QCD. The volume tendency of $I_{s_0}$ is clear, and the
scale-invariant point, i.e., the mobility edge, is easily spotted.

The presence of localised modes in the spectrum of $\Dslash$ in QCD is
by itself not surprising given the formal analogy with random
Hamiltonians. Similarly, it is not surprising that critical properties
at the mobility edge match those of the 3D unitary Anderson model
Hamiltonian~\cite{Giordano:2013taa,Nishigaki:2013uya,Ujfalusi:2015nha},
which has the same dimensionality and symmetry class as $\Dslash$ in
QCD. What \textit{is} surprising though, and calls for an explanation,
is the sudden appearance of localised modes near the origin at
sufficiently high temperatures.

\section{Sea/islands picture}
\label{sec:si}

A qualitative explanation of localisation in QCD was proposed in
Ref.~\cite{Bruckmann:2011cc} and refined in
Refs.~\cite{Giordano:2015vla,Giordano:2016cjs,Giordano:2016vhx}. This
is based on the observation that the local Polyakov loop acts in
practice as a space-dependent effective temporal boundary condition
for the Dirac eigenmodes, modifying the ``twist'' on the wave
functions coming from the antiperiodic boundary conditions.  At high
temperature in the deconfined phase the Polyakov loops get ordered,
and so doing they induce strong correlations in the temporal
direction. In a first approximation one can then imagine the typical
gauge configurations as fluctuating around a strongly ordered
configuration with identical Polyakov loops
$P(\vec{x}\mkern 2mu)=\textrm{diag}(e^{i\phi_a})$,
$\phi_a\in (-\pi,\pi]$, and trivial spatial gauge fields. For this
kind of configuration the eigenmodes and the spectrum can be computed
exactly: eigenmodes are delocalised plane waves, and the spectrum has
a gap $\omega(\phi_{\textrm{PL}})= (\pi- |\phi_{\textrm{PL}}|)T$
determined by the Polyakov-loop phase $\phi_{\textrm{PL}}$ the
farthest away from zero - the larger the distance, the smaller the
gap, as well as the effective temporal twist on the wave functions.
Dynamical fermions then select the trivial Polyakov loop, for which
the twist, the gap, and so the fermion determinant are maximal.  On
typical configurations, the ``sea'' of ordered Polyakov loops contains
``islands'' of fluctuations, where the Polyakov-loop phases are
nontrivial and so reduce the effective temporal twist on the
fermions. Modes supported on these islands are expected to have
eigenvalues inside the gap, at the price of being localised
there. This leads to a relatively low density of localised modes
filling the gap, with a mobility edge developing to separate them from
the delocalised bulk modes [Fig.~\ref{fig:iulsd} (right)].

The most attractive feature of this ``sea/islands'' picture is that it
requires only the ordering of the Polyakov loop to be applicable.
This leads to predict localisation of the low Dirac modes in the
deconfined phase of a generic gauge theory. Moreover, one expects
localised modes to appear right at the critical temperature when the
transition is genuine. Such predictions have been confirmed so far by
all the numerical studies carried out on the lattice.  These include,
besides QCD, also pure gauge theories with group SU(2) in
3+1D~\cite{Kovacs:2010wx}, SU(3) in
3+1D~\cite{Kovacs:2017uiz,Vig:2020pgq} and
2+1D~\cite{Giordano:2019pvc}, $\mathbb{Z}_2$~\cite{Baranka:2021san}
and $\mathbb{Z}_3$~\cite{Baranka:2022dib} in 2+1D, SU(3) with trace
deformation in 3+1D~\cite{Bonati:2020lal}, and a toy model for QCD
with $N_f=3$ fermions on coarse
lattices~\cite{Giordano:2016nuu}. Localised low modes were not found
in the confined phase in any of these cases.

The sea/islands picture as presented above turns out to be too
simplistic, and to require a refinement involving a more prominent
role for spatial gauge fields~\cite{Baranka:2022dib}. Nonetheless,
Polyakov-loop fluctuations away from order still retain their role as
favourable localisation centres also in the refined version of the
picture. Considerations based on Polyakov loops are then expected to
correctly guide our expectations concerning the localisation of low
Dirac modes in the deconfined phase of a gauge theory in most of the
cases.

\section{QCD at finite imaginary chemical potential}
\label{sec:RW}

The introduction of an imaginary chemical potential $\mu_I$ for the
quarks boils down in practice to a modification of the temporal
boundary conditions from antiperiodic to
$\psi(1/T,\vec{x}\mkern 2mu) =
e^{i(\pi+\hat{\mu}_I)}\psi(0,\vec{x}\mkern 2mu)$. This is the same
effect caused by a centre transformation, that modifies all Polyakov
loops to $P(\vec{x}\mkern 2mu)\to z P(\vec{x}\mkern 2mu)$, where
$z=e^{i\zeta}$ is an element of the gauge group centre, and leads to
$\psi(1/T,\vec{x}\mkern 2mu) = e^{i(\pi+\zeta)}\psi(0,\vec{x}\mkern
2mu)$.  Since a centre transformation does not change the gauge
action, this leads to the well-known Roberge--Weiss symmetry of the
partition function $Z(\hat{\mu}_I)$, i.e., the invariance of
$Z(\hat{\mu}_I)$ under
$\hat{\mu}_I\to \hat{\mu}_I + \zeta$~\cite{Roberge:1986mm}. For QCD
where the gauge group is $\textrm{SU}(3)$, with centre $\mathbb{Z}_3$,
this means that $Z(\hat{\mu}_I)$ is periodic with period
$\f{2\pi}{3}$. This periodicity, however, is realised differently at
low and high temperature: at low $T$ in the confined phase,
$Z(\hat{\mu}_I)$ is a smooth periodic function, while at high $T$ in
the deconfined phase it displays lines of first-order phase
transitions at $\hat{\mu}_I= \f{\pi}{3}+\f{2\pi}{3}n$,
$n\in\mathbb{Z}$. At these values of $\hat{\mu}_I$ two of the centre
sectors are equally favoured by fermions and the system has an exact
$\mathbb{Z}_2$ centre symmetry, which at high temperature is
spontaneously broken. The first-order lines terminate at second-order
points in the Ising class at the Roberge-Weiss temperature
$T_{\textrm{RW}}=208(5)~\textrm{MeV}$~\cite{Bonati:2016pwz}.

\begin{figure}[t]
  \centering
  \includegraphics[width=0.5\textwidth]{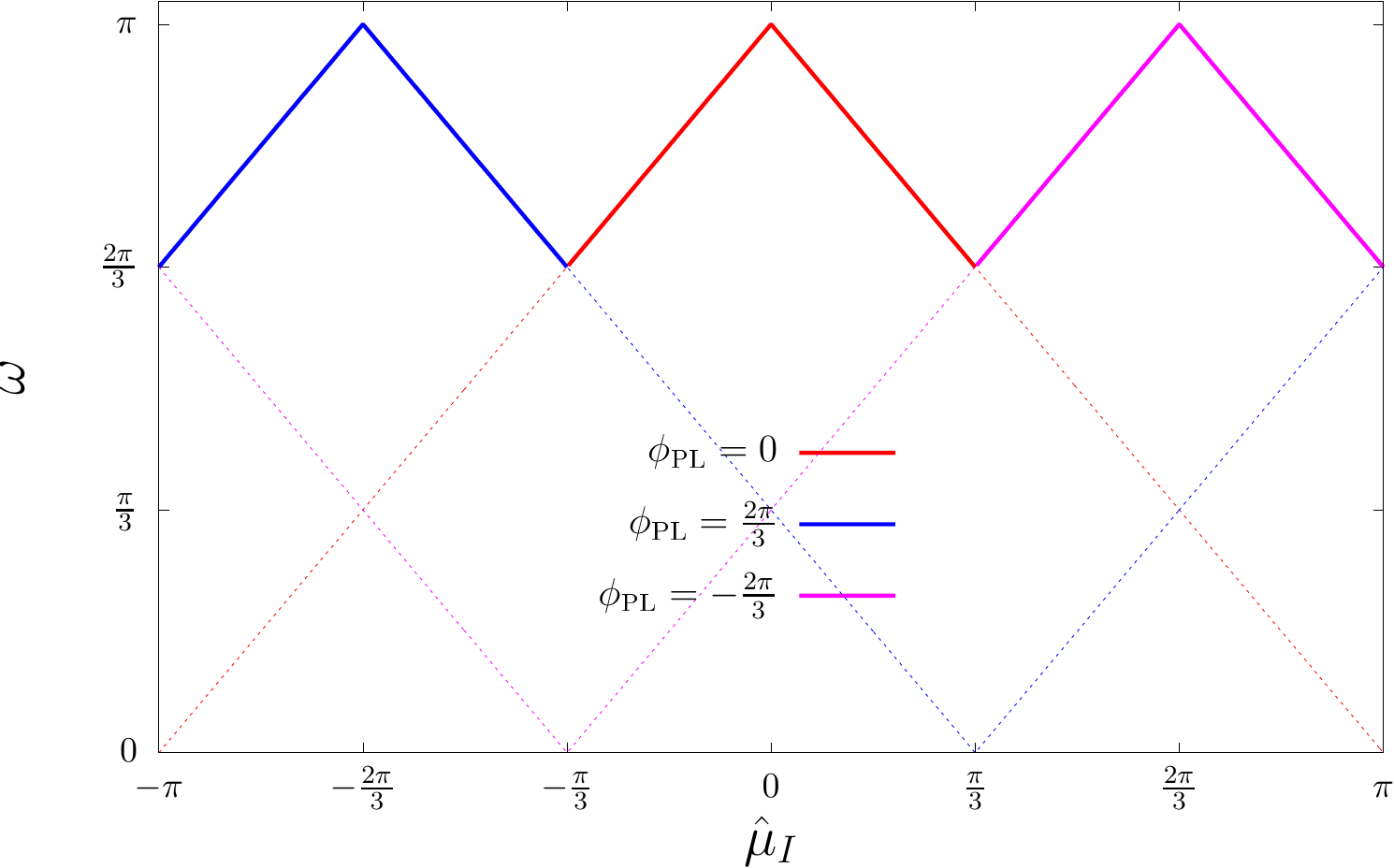} \hfil
  \includegraphics[width=0.47\textwidth]{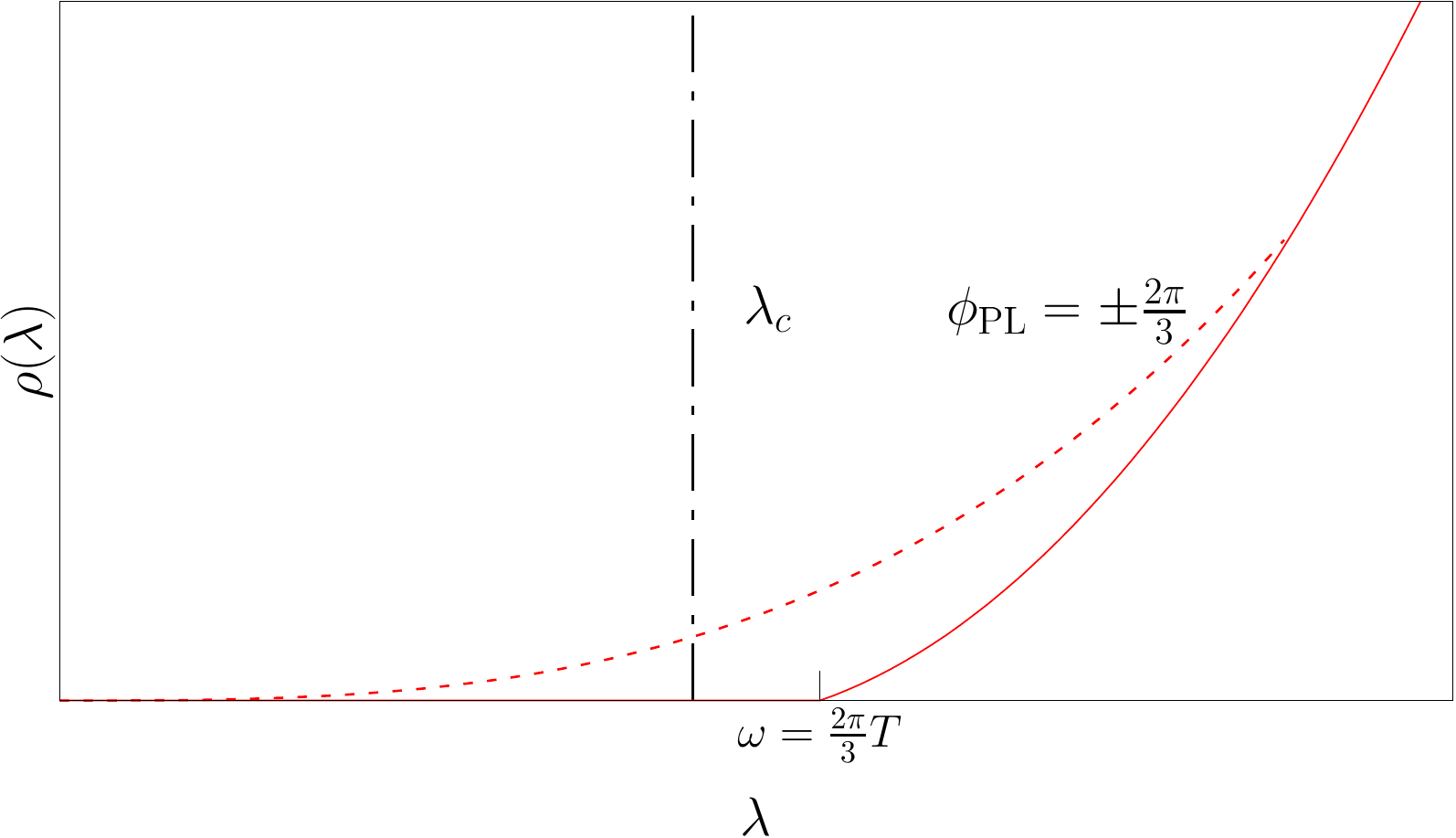}

  \caption{Left: gap function $\omega(\phi_{\textrm{PL}},\hat{\mu}_I)$
    at finite imaginary chemical potential, for $\phi_{\textrm{PL}}$
    corresponding to the three centre elements of
    $\textrm{SU}(3)$. Solid lines indicate the vacuum selected by
    fermions.  Right: sketch of the spectral density at
    $\hat{\mu}_I=\pi$. The solid line is the spectral density in the
    case of vanishing gauge fields, starting at the gap
    $\omega=\f{2\pi}{3}T$, while the dashed line qualitatively shows
    the spectral density after the inclusion of localised modes
    penetrating the gap, and $\lambda_c$ indicates the position of the
    mobility edge.}
\label{fig:si_mu}
\end{figure}

The sea/islands picture discussed in Sec.~\ref{sec:si} applies also at
finite $\mu_I$ after suitably modifying the gap function to
$\omega=\omega(\phi_{\textrm{PL}},\hat{\mu}_I)$. Its behaviour as a
function of $\hat{\mu}_I$ is shown in Fig.~\ref{fig:si_mu} (left) in
the case of Polyakov loops ordered along one of the centre elements of
$\textrm{SU}(3)$. In the case of $\hat{\mu}_I=\pi$ (corresponding to
effectively periodic boundary conditions) it reads
$\omega(\phi_{\textrm{PL}},\pi)=|\phi_{\textrm{PL}}|T$. This shows
that the complex sectors $z = e^{\pm i\f{2\pi}{3}}$ are equally
favoured over the real sector by dynamical fermions, in agreement with
the residual $\mathbb{Z}_2$ centre symmetry at $\hat{\mu}_I=\pi$.
Independently of what sector is chosen when the symmetry breaks
spontaneously, Polyakov-loop fluctuations to the real sector are
favourable localisation centres for the Dirac eigenmodes. In fact, the
gap function is locally vanishing there, $\omega(0,\pi)=0$ [see
Fig.~\ref{fig:si_mu} (left)], and so they can support localised modes
with eigenvalues below the gap
$\omega(\pm\f{2\pi}{3},\pi)=\f{2\pi}{3}T$. The resulting spectral
density is sketched in Fig.~\ref{fig:si_mu} (right). Moving along the
temperature axis at $\hat{\mu}_I=\pi$, one then expects localised
modes to appear right at $T_{\textrm{RW}}$.

\section{Localisation above the Roberge-Weiss temperature}
\label{sec:RW_num}

Studying localisation of low Dirac modes along the $\hat{\mu}_I=\pi$
line provides the rare opportunity of checking the expectations of the
sea/islands picture across a genuine deconfinement phase transition in
the presence of dynamical fermions. This was done by us in
Ref.~\cite{Cardinali:2021fpu} by means of a numerical study on the
lattice. We simulated $N_f=2+1$ QCD at physical values of the quark
masses using 2-stout improved rooted staggered fermions and a
tree-level improved Symanzik gauge action, and studied the statistical
properties of the spectrum of the staggered discretisation of the
Dirac operator, in order to infer the localisation properties of its
eigenmodes. Since the staggered operator is anti-Hermitean, the
discussion of Sec.~\ref{sec:loc} applies without modifications. In
particular, we computed $I_{s_0}(\lambda)$ along the spectrum and
identified the mobility edge $\lambda_c$ as the point where it takes
its critical value, $I_{s_0}(\lambda_c)=I_{s_0}^{\textrm{crit}}$. To
check for discretisation effects we used lattices of temporal
extension $N_t=4,6,8$ in lattice units, and spatial extension $N_s$
fixed by setting the aspect ratio to $N_s/N_t=6$ (as well as 8 for
$N_t=4$).

\begin{figure}[t]
  \centering
  \includegraphics[width=0.4775\textwidth,clip]{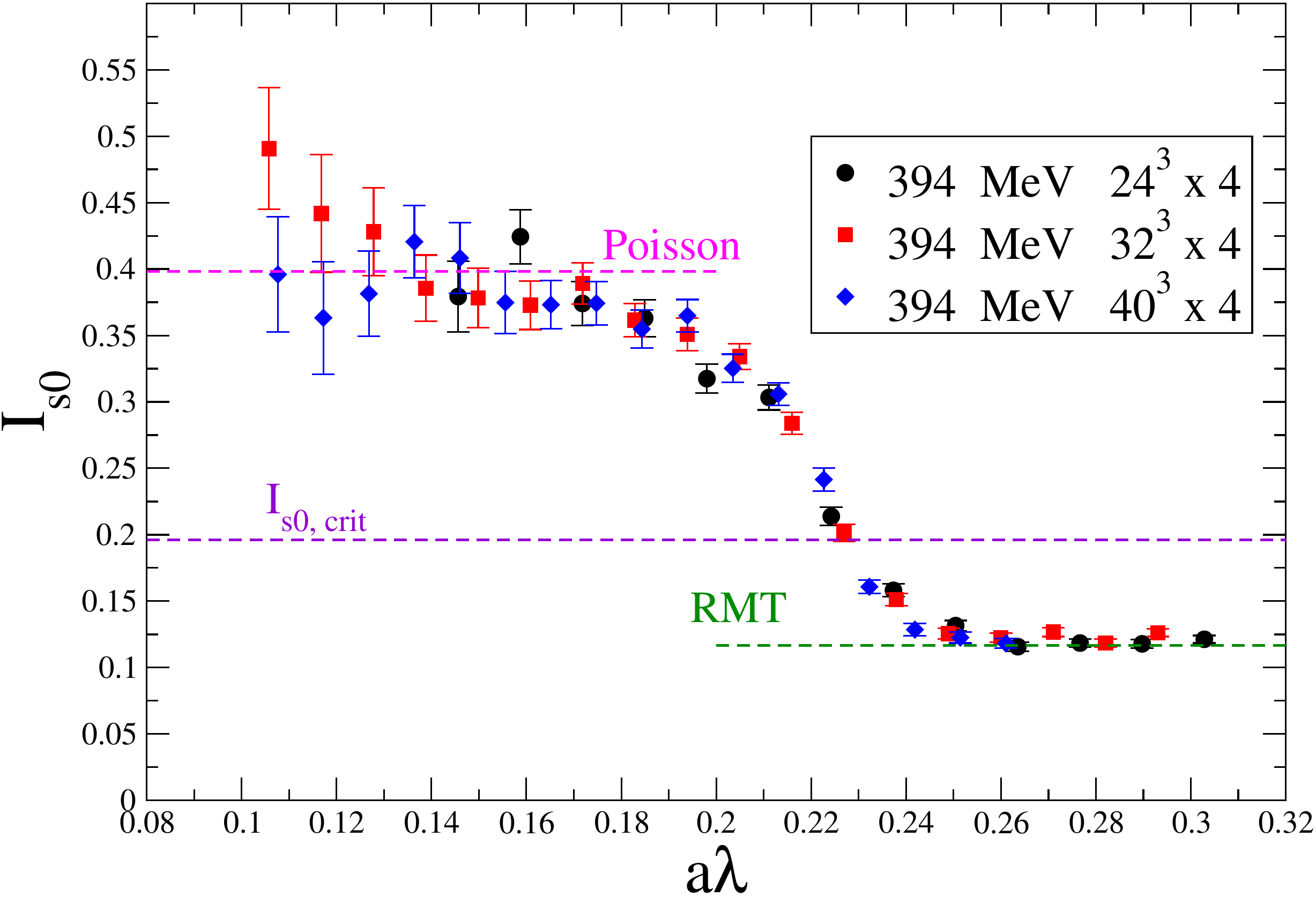}\hfil
  \includegraphics[width=0.4725\textwidth,clip]{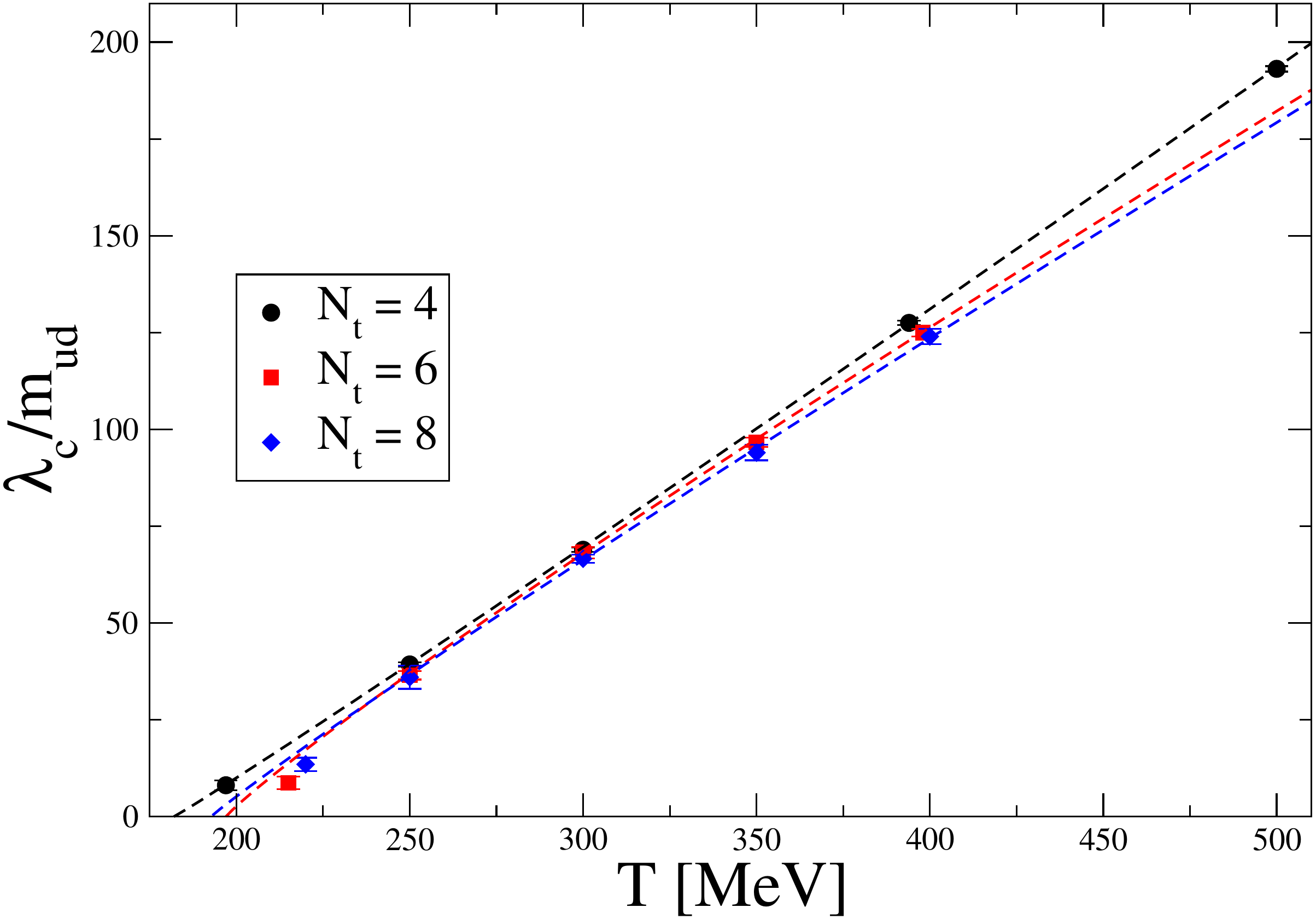}

  \caption{Left: $I_{s_0}$ ($s_0\simeq 0.508$) in QCD at
    $\hat{\mu}_I=\pi$ and $T=394~\textrm{MeV}$ for $N_s^3\times N_t$
    lattices of different spatial size. The Poisson and RMT
    expectation and the critical value
    $I_{s_0}^{\textrm{crit}}$~\cite{Giordano:2013taa} are also
    shown. Right: temperature dependence of the mobility edge in units
    of $m_{\textrm{ud}}$. Figures from Ref.~\cite{Cardinali:2021fpu}.}
\label{fig:RW}       
\end{figure}
\begin{figure}[t]
  \centering
  \includegraphics[width=0.53\textwidth,clip]{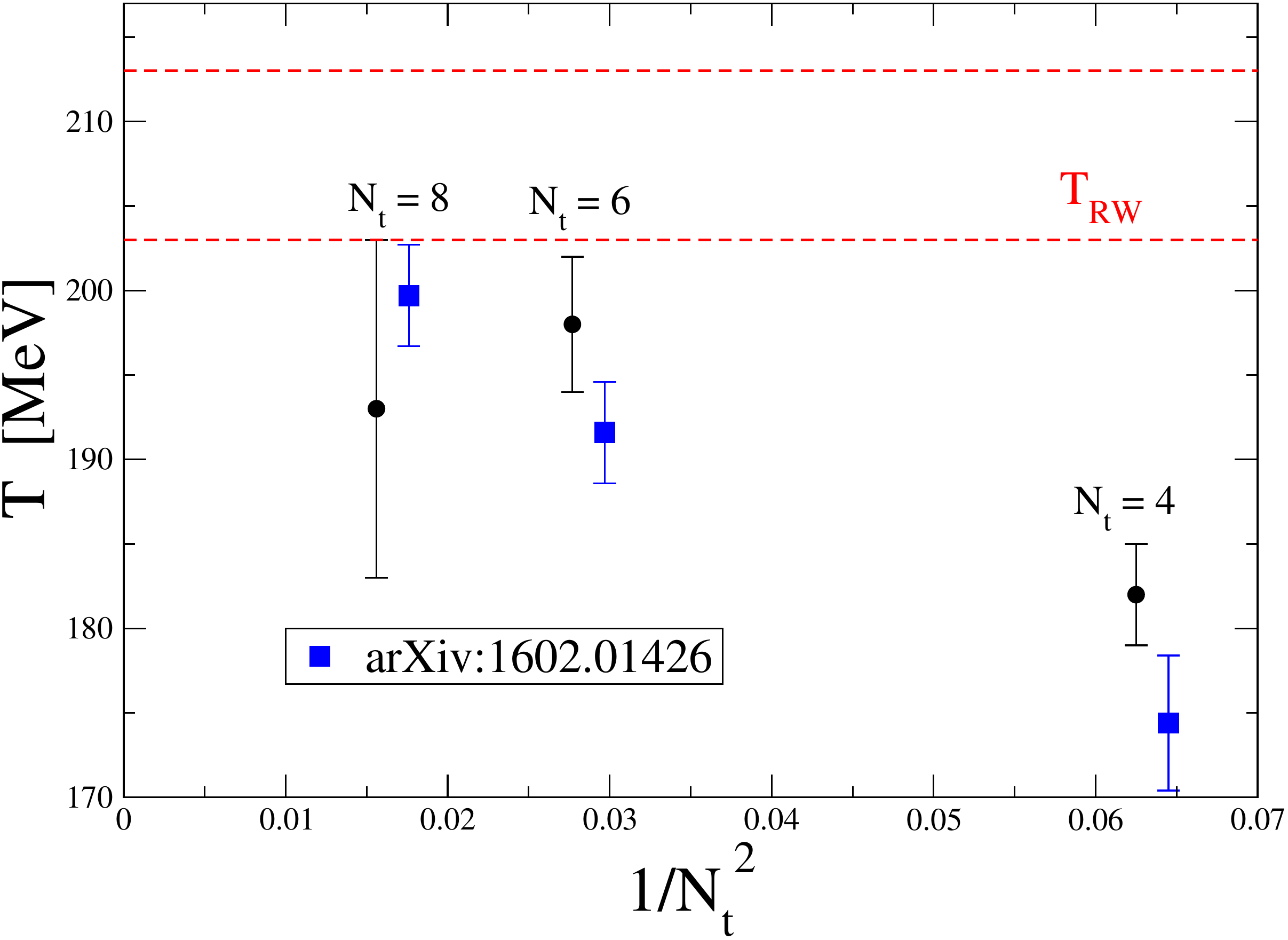}
  \caption{Localisation temperature $T_{\textrm{loc}}(N_t)$ (black
    dots) and Roberge-Weiss temperature $T_{\textrm{RW}}(N_t)$ (blue
    squares; data from Ref.~\cite{Bonati:2016pwz}) for the various
    temporal extensions $N_t$. The error band of the continuum
    extrapolation of $T_{\textrm{RW}}$ is also shown (red dashed
    lines). Figure from Ref.~\cite{Cardinali:2021fpu}.}
\label{fig:RW2}
\end{figure}

An example of our results for $I_{s_0}$ is shown in Fig.~\ref{fig:RW}
(left) for $T=394~\textrm{MeV}$. At this temperature we also checked
for finite-volume effects, obtaining consistent results for
$\lambda_c$ for different $N_s$. The results for $\lambda_c$ in the
whole range of temperatures that we investigated is shown in
Fig.~\ref{fig:RW} (right). There we plot the ratio
$\lambda_c/m_{\textrm{ud}}$, with $m_{\textrm{ud}}$ the mass of the
light quarks, against $T=1/(aN_t)$, for the different choices of $N_t$
(and corresponding lattice spacing $a$). The ratio
$\lambda_c/m_{\textrm{ud}}$ is a renormalisation-group-invariant
quantity~\cite{Kovacs:2012zq,Giordano:2022ghy}, expected to depend
mildly on $a$ and to have a finite continuum limit. This is supported
by our results.

For each set of data at fixed $N_t$ we fitted the temperature
dependence of the mobility edge with a power law,
$ \lambda_c/m_{\textrm{ud}} =
A(N_t)[T-T_{\textrm{loc}}(N_t)]^{B(N_T)}$, with
$T_{\textrm{loc}}(N_t)$ the localisation temperature where the
mobility edge vanishes.\footnote{This is assumed to happen in a
  continuous fashion, as one expects at a second-order phase
  transition where the ordering of Polyakov loops majorly responsible
  for localisation disappears gradually. A jump in the mobility edge
  has been observed at first-order
  transitions~\cite{Bonati:2020lal,Kovacs:2021fwq}.}  Results at the
lowest temperature for $N_t=6,8$ were excluded from the fits, since
they are still affected by large finite-size effects for the available
volumes. Results for $T_{\textrm{loc}}(N_t)$ are shown in
Fig.~\ref{fig:RW2}, together with the corresponding lattice
determinations $T_{\textrm{RW}}(N_t)$ of the Roberge-Weiss temperature
and the resulting continuum extrapolation
$T_{\textrm{RW}}$~\cite{Bonati:2016pwz}. For each $N_t$ the results
for $T_{\textrm{loc}}(N_t)$ and $T_{\textrm{RW}}(N_t)$ are compatible
within errors, which strongly supports that the continuum
extrapolation $T_{\rm loc}$ of the localisation temperature satisfies
$T_{\textrm{loc}}=T_{\textrm{RW}}$.

Our results provide further evidence of the strong connection between
deconfinement and the localisation of the low Dirac modes, that
happens right at the critical temperature, also in the case of a
genuine deconfinement transition in the presence of dynamical
fermions.

\noindent \textit{Acknowledgements} MG was partially supported by the NKFIH grant KKP-126769.

\bibliography{references_RW}

\end{document}